\documentclass[conference]{IEEEtran}
\ifCLASSINFOpdf
\else
\fi
\usepackage[flushleft]{threeparttable}
\usepackage{times}
\usepackage{helvet}
\usepackage{courier}
\makeatletter
\newcommand*{\rom}[1]{\expandafter\@slowromancap\romannumeral #1@}
\makeatother

\usepackage{mathtools}
\usepackage{amssymb}
\usepackage{algorithm,algpseudocode}
\usepackage{multirow}

\usepackage{url}  
\usepackage{graphicx}  

\usepackage[flushleft]{threeparttable}
\usepackage{bm}
\usepackage{array}
\usepackage{hhline}
\usepackage{pgfplots}
\usepackage{tikz}
\usepackage{subcaption}

\usepackage{booktabs}
\usepackage{colortbl}
\usepackage{multirow}

\usepackage{fancyhdr}
\usepackage{kantlipsum}
\fancypagestyle{plain}{
	\fancyhf{}
	\fancyhead[C]{Conference on \LaTeX} 
	
}
\usepackage{eso-pic}
\begin{document}
	\AddToShipoutPictureBG*{
		\AtPageUpperLeft{
			\setlength\unitlength{1in}
			\hspace*{\dimexpr0.5\paperwidth\relax}
			\makebox(0,-0.75)[c]{\textbf{2018 IEEE/ACM International Conference on Advances in Social Networks Analysis and Mining (ASONAM)}}}}

%
\title{Detecting Antagonistic and Allied Communities on Social Media}


\author{\IEEEauthorblockN{Amin Salehi, Hasan Davulcu}
\IEEEauthorblockA{Computer Science and Engineering\\
Arizona State University, Tempe, USA\\
\{asalehi1, hdavulcu\}@asu.edu}}


%


\maketitle

\IEEEoverridecommandlockouts 
\IEEEpubid{\parbox{\columnwidth}{\vspace{8pt}
		\makebox[\columnwidth][t]{IEEE/ACM ASONAM 2018, August 28-31, 2018, Barcelona, Spain}
		\makebox[\columnwidth][t]{978-1-5386-6051-5/18/\$31.00~\copyright\space2018 IEEE} \hfill} \hspace{\columnsep}\makebox[\columnwidth]{}}
\IEEEpubidadjcol

\begin{abstract}
	Community detection on social media has attracted considerable attention for many years. However, existing methods do not reveal the relations between communities. Communities can form alliances or engage in antagonisms due to various factors, e.g., shared or conflicting goals and values. Uncovering such relations can provide better insights to understand communities and the structure of social media. According to social science findings, the attitudes that members from different communities express towards each other are largely shaped by their community membership. Hence, we hypothesize that inter-community attitudes expressed among users in social media have the potential to reflect their inter-community relations. Therefore, we first validate this hypothesis in the context of social media. Then, inspired by the hypothesis, we develop a framework to detect communities and their relations by jointly modeling users' attitudes and social interactions. We present experimental results using three real-world social media datasets to demonstrate the efficacy of our framework.
	
\end{abstract}


%
\IEEEpeerreviewmaketitle

\section{Introduction}
Although community detection plays an important role in providing insights into the structure and function of social media \cite{papadopoulos2012community}, existing community detection methods do not reveal inter-community relations, which are indispensable to deepen our insights. Moreover, to better understand communities, there is a need to uncover their relations. Indeed, social scientists suggest that ``the understanding of policies and practices prevailing within groups will be inadequate unless relations among them are brought into the picture" \cite{sherif1953groups}. A community, or group in social sciences, is defined as a set of users with many intra-group social interactions and few inter-group ones \cite{girvan2001community}, who tend to have mainly positive attitudes towards each other \cite{festinger1950social,lott1965group}.

Several methods \cite{chu2016finding,gao2016detecting,lo2013mining,lo2011mining,zhang2010mining,zhang2013mining} have been proposed to detect antagonistic communities. There are generally two categories of such methods: (1) those which detect antagonistic communities from signed networks \cite{chu2016finding,gao2016detecting,lo2013mining,lo2011mining}, and (2) those which mine antagonistic communities by finding frequent patterns in users' ratings \cite{zhang2010mining,zhang2013mining}. However, these methods suffer from  two main limitations. First, they cannot be applied to a majority of popular social network platforms (e.g., Facebook and Twitter) since these platforms do not provide signed links or users' ratings explicitly. Second, inter-community relations are not restricted to antagonisms. Indeed, communities can also form alliances. 

According to social science findings, inter-community attitudes that individuals express towards each other are largely shaped by their community membership rather than their characteristics or personal relationships \cite{tajfel1979human,billig1973social}. Moreover, Tajfel \cite{tajfel2010social} observed a pair of characteristics in inter-community behavior. First, the members of a community display uniformity in their behavior and attitude towards any other community. Second, they tend to perceive the characteristics and behavior of the members of any other community as undifferentiated. 
Moreover, social scientists suggest that ``the social psychology of intergroup relations is concerned with intergroup behaviour and attitudes" \cite{tajfel2010social}. According to these observations, inter-community attitudes that users express towards each other in social media have the potential to reflect inter-community relations.

In this paper, we propose a framework, namely DAAC, which detects communities and their relations (i.e., antagonism, alliance, or neither) by exploiting users' social interactions (e.g., retweets) and attitudes expressed on social media. Our main contributions are:

\begin{itemize}
	\item Validating the hypothesis suggesting that inter-community attitudes that users express towards each other in social media can reflect the relations of their communities;
	\item Achieving higher performance in detecting communities compared to several standard community detection methods;
	\item Uncovering inter-community relations, i.e., antagonism, alliance, or no relation.
	
\end{itemize}

The rest of the paper is organized as follows. In Section \ref{related_work}, we review related work. In Section \ref{problem_statement}, we formally define the problem of detecting communities and their relations on social media. Section \ref{data-desc} describes  three real-world social media datasets used in our experiments. In Section \ref{proposed_framework_section}, we first validate the aforementioned hypothesis and then present our framework. In Section \ref{experiments}, we demonstrate the effectiveness of the proposed framework. Section \ref{conclusion} concludes the paper and discusses future work.

\section{Related Work}
\label{related_work}
There has been a lot of efforts to detect communities efficiently and accurately. To this end, a wide variety of approaches have been utilized. Modularity-based methods are among the most well-know techniques to detect communities. The modularity measure proposed in \cite{newman2004finding} evaluates whether a division is good enough to form communities. Many variants of modularity-based community detection \cite{clauset2004finding,blondel2008fast} have been developed. Another well-known category includes spectral algorithms \cite{dhillon2004kernel,ding2005equivalence,newman2006finding, salehi2018sentiment} which aims to divide the network into several communities in which most of the interactions are within communities while the number of interactions across communities are minimized. Probabilist approaches \cite{yu2005soft}, in which users are assigned to clusters in a probabilistic way, are also applied to the problem of community discovery. There are a variety of approaches such as information theory based methods \cite{rosvall2008maps}, random walk techniques \cite{harel2001clustering,pons2006computing}, and model-based methods \cite{raghavan2007near,gregory2010finding} to tackle this problem.

Although there has been a great deal of efforts to detect communities, to the best of our knowledge, no previous work has been proposed to uncover the existence of antagonism and alliance between communities. However, some efforts have been made \cite{chu2016finding,gao2016detecting,lo2013mining,lo2011mining,zhang2010mining,zhang2013mining} to detect only antagonistic communities. These methods can be roughly divided into two main categories. First category includes the methods \cite{zhang2010mining,zhang2013mining} utilizing frequent patterns in users' ratings to mine antagonistic communities. Second category includes the methods \cite{chu2016finding,gao2016detecting,lo2013mining,lo2011mining} utilizing signed networks, having trust and distrust links, to detect antagonistic communities. A majority of these methods \cite{gao2016detecting,lo2013mining,lo2011mining} detect a pair of subgraphs with most trust links preserved between the members of each subgraph and most distrust links remained between the members of different subgraphs. These methods are limited to detecting only a pair of antagonistic communities. To address this limitation, another method \cite{chu2016finding} has been proposed to detect multiple antagonistic communities by finding several dense subgraphs with the mentioned property. However, as experiments in \cite{chu2016finding} show such methods usually end up with large number of small subgraphs due to high sparsity of users' interactions in social media.


\section{Problem Statement}
\label{problem_statement}
We first begin with the introduction of the notations used in the paper as summarized in Table \ref{tab:notations}. Let $\mathcal{U}=\{u_1,u_2,...,u_n\}$ be the set of $n$ users and $\mathcal{C}=\{c_1,c_2,...,c_k\}$ indicate the set of $k$ communities. $\mathbf{R}\in \mathbb{R}_+^{n \times n}$ denotes the  social interaction matrix, where $\mathbf{R}_{i,j}$ corresponds to the number of social interactions between user $u_i$ and user $u_j$. $\mathbf{S}\in \mathbb{R}^{n \times n}$ indicates the attitude matrix, where the positive/negative value of $\mathbf{S}_{i,j}$ corresponds to the positive/negative attitude strength of user $u_i$ towards user $u_j$. $\mathbf{U}\in \mathbb{R}_+^{n \times k}$ indicates the community membership matrix, in which $\mathbf{U}_{i,l}$ corresponds to the membership strength of user $u_i$ to community $c_l$. $\mathbf{H}\in \mathbb{R}^{k \times k}$ denotes intra/inter-community relation matrix, where $\mathbf{H}_{i,j}$, if $i\neq j$, corresponds to the strength and type of inter-community relation between community $c_i$ and community $c_j$; the negative, positive, and zero value of $\mathbf{H}_{i,j}$ indicates antagonism, alliance, or no relation between community $c_i$ and community $c_j$, respectively. Moreover, $\mathbf{H}_{i,i}$ corresponds to the intra-community attitudes that the members of community $c_i$ have expressed towards each other. We define the symmetric normalization of $\mathbf{R}$ as $\overset{\sim}{\mathbf{R}}=\mathbf{D}^{-1/2} \mathbf{R} \mathbf{D}^{-1/2}$, where $\mathbf{D}=diag(d_1,d_2,...,d_n)$ is the degree matrix of $\mathbf{R}$ and the degree of user $u_i$ is $d_i=\sum_{j=1}^{n}\mathbf{R}_{i,j}$. We separate positive and negative parts of matrix $\mathbf{A}$ as $\mathbf{A}_{i,j}^+=(|\mathbf{A}_{i,j}|+\mathbf{A}_{i,j})/2$ and $\mathbf{A}_{i,j}^-=(|\mathbf{A}_{i,j}|-\mathbf{A}_{i,j})/2$.

By using the aforementioned notations, the problem of detecting communities and their relations on social media can be defined as: \textit{Given  social interaction matrix $\mathbf{R}$ and  attitude matrix $\mathbf{S}$, we aim to obtain community membership matrix $\mathbf{U}$ and intra/inter-community relation matrix $\mathbf{H}$.}

\begin{table}
	\centering
	\caption{Notations used in the paper}
	\label{tab:notations}
	\small
	\begin{tabular}{|c|l|} \hline
		Notation & Explanation\\ \hline
		$\mathcal{U}$ & The set of users\\ \hline
		$\mathcal{C}$ & The set of communities\\ \hline
		$n$ & The number of users\\ \hline
		$k$ & The number of communities\\ \hline
		$\mathbf{R}$ & The  social interaction matrix\\ \hline
		$\mathbf{S}$ & The  attitude matrix\\ \hline
		$\mathbf{U}$ & The community membership matrix\\ \hline
		$\mathbf{H}$ & The community intra/inter-relation matrix\\ \hline
		$\overset{\sim}{\mathbf{R}}$ & Symmetrically normalized matrix $\mathbf{R}$\\ \hline
		$\mathbf{D}$ & Degree matrix of $\mathbf{R}$\\ \hline
		$\mathbf{A}^+$ & The positive part of matrix $\mathbf{A}$ (i.e., $(|\mathbf{A}|+\mathbf{A})/2$) \\ \hline
		$\mathbf{A}^-$ & The negative part of matrix $\mathbf{A}$ (i.e., $(|\mathbf{A}|-\mathbf{A})/2$) \\ \hline
	\end{tabular}
\end{table}

\section{Data Description}
\label{data-desc}
Politics is a domain in which it is common among political parties (i.e., communities) to form alliances or engage in antagonisms. To validate the aforementioned hypothesis and evaluate our proposed framework, we use the following political Twitter datasets:
\begin{itemize}
	\item \textbf{US Dataset} consists of the tweets posted by 583 politicians from two major US political parties (the Republican Party and the Democratic Party) from August 26 to November 29, 2016.  For the period of time that this dataset covers, there were antagonisms between these parties particularly due to the 2016 presidential election campaigning \cite{lilleker2016us}.
	\item \textbf{Australia Dataset} consists of the tweets posted by 225 user accounts, including politicians and political groups, from five major Australian political parties (the Liberal Party, the National Party, the Liberal National Party, the Greens, and the Labor Party) from January 1 to November 18, 2016. For several decades, there has been a coalition among the Liberal Party, the National Party, and the Liberal National Party \cite{clune2016contemporary}. In the 2016 federal election, all relations between the parties were antagonistic except the relations between the members of the coalition,.
	\item \textbf{UK Dataset} consists of the tweets posted by 389 user accounts, including politicians and political groups, from five major UK political parties (the Conservative Party, the Labour Party, the Scottish National Party, the Liberal Democrats Party, and the UK Independence Party) from January 1 to October 31, 2015. There were antagonism among five major UK political parties in this period of time, especially due to the 2015 general election campaigning \cite{moran2015politics}.
\end{itemize}

\textbf{Preprocessing}: For all datasets, we remove the users who do not have any retweet (i.e., social interaction). Table \ref{tab:statistics} shows the statistics of the preprocessed datasets. All users in the datasets have been labeled with their corresponding parties, and these labels are used to evaluate our proposed method. 



Although aspect-based sentiment classification techniques \cite{pontiki2016semeval} have been proposed to capture users' attitudes towards entities, publicly available training datasets are either too small or domain-oriented, making such techniques incapable to tackle real-world problems. Therefore, we use the following technique to extract the attitudes that users express towards each other in social media. Given each message in which author $u_i$ has mentioned user $u_j$, we add the strength of the message's sentiment to the corresponding elements of matrix $\mathbf{S}$ (i.e., $\mathbf{S}_{i,j}$). 
Even though some messages may carry a negative sentiment, the author may not necessarily have an antagonistic attitude towards a mentioned user. To alleviate this problem, we ignore such messages if there is a social interaction (i.e., retweet) between the author and the mentioned user since a social interaction indicates the presence of a good relationship \cite{conover2011political}. We utilize SentiStrength \cite{thelwall2010sentiment} to detect the sentiment polarity and strength of messages. We have made the code and datasets used in this paper available\footnote{\url{https://github.com/amin-salehi/DAAC}}.


\begin{table}
	\centering
	\caption{The statistics of the cleaned datasets.}
	\label{tab:statistics}
	\begin{tabular}{@{} l c c c c @{}}
		\toprule
		& US & Australia & UK \\ \hline
		\# of tweets & 111,743 & 159,499 & 267,085 \\
		\# of retweets & 17,724 & 21,111 & 14,892 \\
		\# of mentions & 8,470 & 14,996 & 33,462 \\
		\# of user accounts & 583 & 225 & 389\\
		\# of true communities & 2 & 5 & 5\\
		\# of allied relations & 0 & 3 & 0 \\
		\# of antagonistic relations & 1 & 7 & 10 \\ \bottomrule
	\end{tabular}
\end{table}

\section{The Proposed Framework}
\label{proposed_framework_section}
In this section, we first demonstrate the existence of a significant level of correlation between the type of inter-community relation (i.e., alliance or antagonism) between two communities and the type of sentiment (i.e., positive or negative) that members from these communities expressed towards each other. Next, we propose our framework.


\subsection{Validating the Hypothesis}
According to social science findings \cite{tajfel1979human,billig1973social}, the attitudes that members from different communities express towards each other are largely shaped by their community membership. Therefore, we hypothesize that inter-community attitudes expressed among users towards each other in social media have the potential to reflect inter-community relations. However, the findings borrowed from social sciences do not necessarily hold in social media due to many factors, such as the validity and representativeness of available information \cite{tufekci2014big,ruths2014social}. Moreover, the attitudes that users express towards each other in social media might result from users' personal relationships. Therefore, in this section, we aim to verify our hypothesis by answering the following two questions. With this respect, we utilize the Australia dataset since it is the only dataset containing both allied and antagonistic relations.


\begin{itemize}
	\item Are the communities of two users who express negative attitudes towards each other more likely to be in antagonism?
	\item Are the communities of two users who express positive attitudes towards each other more likely to be in alliance?
\end{itemize}

We first answer the former by using the following procedure inspired by \cite{beigi2016signed}. For each pair of users $(u_i, u_j)$ who are from different communities and have expressed negative attitudes towards each other (i.e., $ \mathbf{S}_{i,j}<0$), we randomly select a user $u_k$ where users $u_i$ and $u_k$ are from different communities and have not expressed negative attitudes towards each other (i.e., $\mathbf{S}_{i,k} \ge 0$). Then, we check whether there is antagonism between the communities of $u_i$ and $u_j$ and between the communities of $u_i$ and $u_k$. If there is antagonism between the communities of $u_i$ and $u_j$, we set $t_p=1$; otherwise $t_p=0$. Similarly, if there is antagonism between the communities of $u_i$ and $u_k$, we set $t_r=1$; otherwise $t_r=0$. Let vector $T_p$ denote the set of all $t_p$s for pairs of users from different communities who have expressed negative attitudes towards each other, and vector $T_r$ denote the set of all $t_r$s for pairs of users from different communities who have not expressed negative attitudes towards each other.

We conduct a two-sample t-test on $T_p$ and $T_r$. The null hypothesis $H_0$ and alternative hypothesis $H_1$ are defined as follows:

\begin{equation}
\label{t-test}
\begin{aligned}
H_0:T_p \le T_r, \quad H_1:T_p > T_r
\end{aligned}
\end{equation}


The null hypothesis is rejected at significance level $a=0.01$ with p-value of $3.56\mathrm{e}{-105}$. Therefore, the result of the two-sample t-test demonstrates that \textit{the communities of two users who express negative attitudes towards each other are highly probable to be in antagonism}. We apply a similar procedure to answer the second question. For brevity, we only report the result of the two-sample t-test. The null hypothesis is rejected at significance level $a=0.01$ with p-value of $1.57\mathrm{e}{-26}$. As a result, we conclude that \textit{the communities of two users who express positive attitudes towards each other are highly probable to be in alliance}.

\subsection{Modeling Users' Attitudes}
In the previous section, we demonstrated that inter-community attitudes expressed by users can reflect the  relation of their communities in the context of social media. Inspired by this observation, we propose a model which uncovers intra/inter-community relations by exploiting the attitudes users express towards each other as,

\begin{equation}
\label{users-attitudes-model}
\begin{aligned}
\underset{\mathbf{U},\mathbf{H}}{\text{min}}\quad & ||\mathbf{W} \odot (\mathbf{S} - \mathbf{U} \mathbf{H} \mathbf{U}^T) ||_F^2 \\
\text{s.t.}\quad & \mathbf{U} \geq 0.
\end{aligned}
\end{equation}

where $\odot$ is Hadamard product, $\mathbf{W}_{i,j}$ controls the contribution of $S_{i,j}$ in the model, and a typical choice of $\mathbf{W}\in \mathbb{R}_+^{n \times n}$ is,
\begin{equation}
\begin{aligned}
\mathbf{W} = \left\{
\begin{array}{ll}
0, \text{ if } \mathbf{S}=0\\
1, \text{ otherwise }\\
\end{array}
\right.
\end{aligned}
\end{equation}

Given communities $c_{i}$ and $c_{j}$, Eq. \eqref{users-attitudes-model}  aims to uncover their inter-community relation $\mathbf{H}_{i,j}$ by using their attitudes. To this end, $\mathbf{U}_{:,i}\mathbf{H}_{i,j}\mathbf{U}_{:,j}^T$ estimates the inter-community attitudes among the members of these two communities as presented in matrix $\mathbf{S}$. Since the non-negativity constraint only holds on $\mathbf{U}$, $\mathbf{H}_{i,j}$ will be negative, positive, or zero if the members of two communities have generally expressed negative, positive, or no attitudes towards each other, respectively. The lower the negative value of $\mathbf{H}_{i,j}$ is, the more antagonistic communities $c_i$ and $c_j$ are. On the other hand, the larger the positive value of $\mathbf{H}_{i,j}$ is, the more allied communities $c_i$ and $c_j$ are. Moreover, $\mathbf{H}_{i,i}$ indicates the intra-community attitudes that the members of community $c_i$ have expressed towards each other.

\subsection{Modeling Social Interactions}
Social interactions are one of the most effective sources of information to detect communities \cite{papadopoulos2012community}. In this section, we aim to cluster users into $k$ communities with the most social interactions within each community and the fewest social interactions between communities. To this end, we use the following model,
\begin{equation}
\label{proposed_graph-reg}
\begin{aligned}
\underset{\mathbf{U}}{\text{max}}\quad & Tr(\mathbf{U}^T \overset{\sim}{\mathbf{R}} \mathbf{U}) \\
\text{s.t.}\quad & \mathbf{U} \geq 0, \mathbf{U}^T \mathbf{U} = \mathbf{I}.
\end{aligned}
\end{equation}

where $\mathbf{I}$ is the identity matrix with the proper size. In fact, Eq. \eqref{proposed_graph-reg} is equivalent to the nonnegative relaxed normalized cut as put forth in \cite{ding2005equivalence}.

\begin{table*}
	\centering
	\small
	\caption{Comparison of community detection methods.}
	\label{tab:performance}
	\begin{tabular}{| l | c c c | c c c | c c c |}
		\hhline{~|*9{-}|}
		\multicolumn{1}{l|}{} & \multicolumn{3}{c|}{US dataset} & \multicolumn{3}{c|}{Australia dataset} & \multicolumn{3}{c|}{UK dataset}  \\
		\hline
		Method & NMI & ARI & Purity & NMI & ARI & Purity & NMI & ARI & Purity \\ \hline \hline
		Louvain & 0.431095 & 0.386347 & 0.943396 & 0.825234 & 0.833025 & 0.942222 & 0.858118 & 0.841718 & 0.987147 \\
		InfoMap & 0.431437 & 0.351938 & 0.946826 & 0.831903 & 0.831737 & 0.942222 & 0.909684 & 0.928716 & \textbf{0.992288} \\
		Leading eigenvectors & 0.580117 & 0.678051 & 0.938250 & 0.779931 & 0.573488 & 0.693334 & 0.913700 & 0.9533703 & 0.982005 \\
		CNM & 0.502925 & 0.487620 & 0.945111 & 0.842446 & 0.848269 & 0.937778 & 0.939093 & 0.971631 & 0.984576 \\
		Label propagation & 0.600777 & 0.655619 & 0.958833 & 0.822237 & 0.826724 & 0.937778 & 0.958379 & 0.979031 & 0.989717 \\
		Soft clustering & 0.735760 & 0.829228 & 0.955403 & 0.841264 & 0.812856 & 0.844444 & 0.951260 & 0.974292 & 0.987147 \\
		DAAC & \textbf{0.768307} & \textbf{0.854484} & \textbf{0.962264} & \textbf{0.903691} & \textbf{0.908264} & \textbf{0.951111} & \textbf{0.958806} & \textbf{0.978770} & 0.989717 \\
		\hline
	\end{tabular}
\end{table*}


\subsection{The Proposed Framework DAAC}

We separately introduced our models to utilize users' attitudes and social interactions. In this section, we propose our framework DAAC, which jointly exploits these two models to uncover communities and their relations. The proposed framework requires solving the following optimization problem,
\begin{equation}
\label{proposed_framework}
\begin{aligned}
\underset{\mathbf{U},\mathbf{H}}{\text{min}}\quad & \bm{\mathcal{F}} = ||\mathbf{W} \odot (\mathbf{S} - \mathbf{U} \mathbf{H} \mathbf{U}^T) ||_F^2 - \lambda Tr(\mathbf{U}^T \overset{\sim}{\mathbf{R}} \mathbf{U})\\
\text{s.t.}\quad & \mathbf{U} \geq 0, \mathbf{U}^T \mathbf{U} = \mathbf{I}.
\end{aligned}
\end{equation}
where $\lambda$ is a non-negative regularization parameter controlling the contribution of social interactions in the final solution.

\begin{algorithm}
	\caption{The Proposed Algorithm for DAAC}\label{alg:1}
	\begin{algorithmic}[1]
		\Statex{\textbf{Input:} attitude matrix $\mathbf{S}$ and social interaction matrix $\mathbf{R}$}
		\Statex{\textbf{Output:} community membership matrix $\mathbf{U}$ and intra/inter-community relation matrix $\mathbf{H}$}
		\State {Initialize $\mathbf{U}$ and $\mathbf{H}$ randomly where $\mathbf{U}\geq 0$}
		\While{ not convergent}
		\State Update $\mathbf{U}$ according to Eq. \eqref{eq:U_updating}
		\State Update $\mathbf{H}$ according to Eq. \eqref{eq:H_updating}
		\EndWhile
		
	\end{algorithmic}
\end{algorithm}

Since the optimization problem in Eq. \eqref{proposed_framework} is not convex with respect to variables $\mathbf{U}$ and $\mathbf{H}$ together, there is no guarantee to find the global optimal solution. As suggested by \cite{lee2001algorithms}, we introduce an alternative scheme to find a local optimal solution of the optimization problem. The key idea is optimizing the objective function with respect to one of the variables $\mathbf{U}$ or $\mathbf{H}$, while fixing the other one. The algorithm keeps updating the variables until convergence.

Optimizing the objective function $\mathcal{F}$  with respect to $\mathbf{U}$ leads to the following update rule,

\begin{equation}
\label{eq:U_updating}
\mathbf{U} = \mathbf{U} \odot \sqrt{\frac{\mathbf{E}_1^+ + \mathbf{E}_2^+ + \mathbf{E}_3^- + \mathbf{E}_4^- + \lambda \overset{\sim}{\mathbf{R}} \mathbf{U} + \mathbf{U}\mathbf{\Gamma}^-} {\mathbf{E}_1^- + \mathbf{E}_2^- + \mathbf{E}_3^+ + \mathbf{E}_4^+ + \mathbf{U} \mathbf{\Gamma}^+}}
\end{equation}
\noindent where,
\begin{align}
& \mathbf{E}_1 = -(\mathbf{W} \odot \mathbf{W} \odot \mathbf{S}) \mathbf{U} \mathbf{H}^T  \\
& \mathbf{E}_2 = -(\mathbf{W} \odot \mathbf{W} \odot \mathbf{S})^T \mathbf{U} \mathbf{H} \\
& \mathbf{E}_3 = (\mathbf{W} \odot \mathbf{W} \odot \mathbf{U} \mathbf{H} \mathbf{U}^T) \mathbf{U} \mathbf{H}^T \\
& \mathbf{E}_4 = (\mathbf{W} \odot \mathbf{W} \odot \mathbf{U} \mathbf{H} \mathbf{U}^T)^T \mathbf{U} \mathbf{H} \\
& \mathbf{\Gamma} \ \ = - \mathbf{U}^T \mathbf{E}_1 - \mathbf{U}^T \mathbf{E}_2 - \mathbf{U}^T \mathbf{E}_3 - \mathbf{U}^T \mathbf{E}_4 \\
& \qquad \  +  \lambda \mathbf{U}^T \overset{\sim}{\mathbf{R}} \mathbf{U}
\end{align}

\noindent The details are given in the Appendix.

\noindent The derivative of $\bm{\mathcal{F}}$ with respect to $\mathbf{H}$ is as follows:

\begin{equation}
\begin{aligned}
\frac{\partial \bm{\mathcal{F}}}{\partial \mathbf{H}} = & -2\mathbf{U}^T (\mathbf{W} \odot \mathbf{W} \odot \mathbf{S}) \mathbf{U} \\
& -2 \mathbf{U}^T (\mathbf{W} \odot \mathbf{W} \odot \mathbf{U} \mathbf{H} \mathbf{U}^T) \mathbf{U}
\end{aligned}
\end{equation}

\noindent Thus, the update rule of $\mathbf{H}$ is as follows:
\begin{equation}
\label{eq:H_updating}
\mathbf{H} = \mathbf{H} - \alpha \frac{\partial \bm{\mathcal{F}}}{\partial \mathbf{H}}
\end{equation}

\noindent where $\alpha$ is the learning rate for updating $\mathbf{H}$.

The detailed algorithm for DAAC is shown in Algorithm \ref{alg:1}. We briefly review Algorithm 1. In line 1, it randomly initializes $\mathbf{U}$ and $\mathbf{H}$. From line 2 to 5, it updates $\mathbf{U}$ and $\mathbf{H}$ until convergence is achieved.

\subsection{Time Complexity}
In Algorithm \ref{alg:1}, the most costly operations are the matrix multiplications in update rules Eq. \eqref{eq:U_updating} and Eq. \eqref{eq:H_updating} on which we focus in this section. $\mathbf{W}$ and $\mathbf{R}$ are usually very sparse matrices, so let $N_w$ and $N_r$ denote the number of non-zero elements of $\mathbf{W}$ and $\mathbf{R}$, respectively. The time complexities of  Eq. \eqref{eq:U_updating} and Eq. \eqref{eq:H_updating} are described as follows:

\begin{itemize}
	\item We first focus on the time complexity of Eq. \eqref{eq:U_updating}. Note that $\mathbf{W} \odot \mathbf{W} \odot \mathbf{S}$ needs to be calculated once. Therefore, the time complexities of both $\mathbf{E_1}$ and $\mathbf{E_2}$ are $\mathcal{O}(N_w k + n k^2)$ thanks to the sparsity of matrices $\mathbf{W}$ and $\mathbf{S}$. The time complexity of $\mathbf{W} \odot \mathbf{W} \odot \mathbf{U} \mathbf{H} \mathbf{U}^T$ is  $\mathcal{O}(N_w n + n k^2 + n^2 k)$. The number of non-zero values of  $\mathbf{W} \odot \mathbf{W} \odot \mathbf{U} \mathbf{H} \mathbf{U}^T$ is the same as $\mathbf{W}$ owing to the sparsity of $\mathbf{W}$. Thus, the time complexities of both $\mathbf{E_3}$ and $\mathbf{E_4}$ are $\mathcal{O}(N_w n + n k^2 + n^2 k)$. Using a similar procedure, the time complexities of $\overset{\sim}{\mathbf{R}} \mathbf{U}$ and $\mathbf{\Gamma}$ are $\mathcal{O}(N_r k)$ and $\mathcal{O}(N_w n + n k^2 + n^2 k + N_r k)$, respectively. As a result, the time complexity of Eq. \eqref{eq:U_updating} is $\mathcal{O}(N_w (n+k) + N_r k + n k^2 + n^2 k)$.
	
	\item Now we provide the time complexity of Eq. \eqref{eq:H_updating}. The cost of $\mathbf{U}^T (\mathbf{W} \odot \mathbf{W} \odot \mathbf{S})$ is $\mathcal{O}(N_w k)$ thanks to the sparsity of $\mathbf{W}$. Thus, the time complexity of $\mathbf{U}^T (\mathbf{W} \odot \mathbf{W} \odot \mathbf{S}) \mathbf{U}$ is $\mathcal{O}(N_w k + n k^2)$. Similarly, the cost of $ \mathbf{U}^T (\mathbf{W} \odot \mathbf{W} \odot \mathbf{U} \mathbf{H} \mathbf{U}^T) \mathbf{U}$ is $\mathcal{O}(N_w n + n k^2 + n^2 k)$. Therefore, the time complexity of Eq. \eqref{eq:H_updating} is $\mathcal{O}(N_w (n+k) + n k^2 + n^2 k)$.
\end{itemize}

Hence, the time complexity of Algorithm \ref{alg:1} is $\mathcal{O}(i (N_w (n+k)+ N_r k + n k^2 + n^2 k))$ where $i$ is the number of iterations required for the convergence. Our framework can be applied to large scale social network platforms by exploiting distributed approaches outlined in \cite{liu2010distributed,gemulla2011large,li2014fast}.

\section{Experiments}
\label{experiments}
To evaluate our proposed framework, we design the required experiments to answer the following two questions.
\begin{enumerate}
	\item How effective is the proposed framework compared to the standard community detection methods?
	\item How effective is our framework in discovering inter-community relations?
\end{enumerate}

In the next section, we first compare the performance of several well-known community detection methods with DAAC. Then, we evaluate the effectiveness of our framework in uncovering inter-community relations. Finally, we study the sensitivity of our framework with respect to regularization parameter $\lambda$. For the experiments, we set the number of communities for any method, if it is required, as the true number of communities (i.e., parties) in each dataset.

\subsection{Evaluation of Community Detection}

\subsubsection{Baselines}
In order to demonstrate the efficacy of DAAC, we compare it with six well-known community detection methods presented as follows:
\begin{itemize}
	\item \textbf{Louvain:} This method \cite{blondel2008fast} greedily maximizes the benefit function known as modularity to detect communities.
	\item \textbf {InfoMap:} This baseline \cite{rosvall2008maps} is based on information theory and compresses the description of random walks in order to find communities.
	\item \textbf{Leading eigenvectors:} Newton  \cite{newman2006finding} presents a formulation of modularity in a matrix form, namely modularity matrix. Then, he proposes to use the eigenvectors of modularity matrix to detect communities.
	\item \textbf{CNM:} This method \cite{clauset2004finding} uses a greedy approach to find the divisions of the network which maximizes the modularity.
	\item \textbf{Label propagation:} \cite{raghavan2007near}
	This method initially assigns unique labels to users. Then, in each iteration, users adopt the label that most of their neighbors posses. Finally, users with the same label fall into the same community. 
	\item \textbf{Soft clustering:} This baseline \cite{yu2005soft} assigns users to communities in a probabilistic way.
\end{itemize}

\begin{table}
	\centering
	\small
	\caption{The uncovered relations between detected communities (i.e., parties) by using DAAC in the US dataset.}
	\label{tab:us_relations}
	\begin{tabular}{|m{1.5cm}|m{1.5cm}|m{1.5cm}|@{}m{0cm}@{}}
		\hhline{~|*3{-}|}
		\multicolumn{1}{c|}{} & Republicans & Democrats &\tabularnewline[2ex]  \hhline{|*3{-}|}\noalign{\vskip.2pt}
		Republicans & \cellcolor{green!40} \hspace{6mm} 259 & \cellcolor{red!40} \hspace{6mm}-138 & \tabularnewline[2ex] \hhline{|*3{-}|}\noalign{\vskip.2pt}
		Democrats & \cellcolor{red!40} \hspace{6mm}-138 & \cellcolor{green!40} \hspace{6mm}112 & \tabularnewline[2ex] \hhline{|*3{-}|}\noalign{\vskip.2pt}
	\end{tabular}
	\\
	\emph{Note:} all values in the table are rounded.
\end{table}

\subsubsection{Performance Measures}
To evaluate the performance of the methods, we utilize three following measures which are frequently used for community detection evaluation: Normalized Mutual Information (NMI), Adjusted Rand Index (ARI), and Purity.

\subsubsection{Experimental Results}
We run all methods with their hyperparameters initialized from $\{10^x | x \in [0,9]\}$. Table \ref{tab:performance} shows the best result for each method. According to the table, we can make the following observations:
\begin{itemize}
	\item Our proposed framework achieves the highest performance in terms of NMI and ARI for all three datasets. In terms of Purity, it also achieves the best in US and Australia datasets. In the UK dataset, only InfoMap obtains higher Purity compared to our framework since it generates a large number of communities (e.g., 11 communities for the UK dataset) for sparse graphs such as social media networks.
	\item Our framework achieves its highest performance with large values of regularization parameter $\lambda$ (e.g., $10^7$). This implies that social interactions are more effective in detecting communities compared to users' attitudes.  We will study more on the impact of the regularization parameter in Section \ref{reg-parameter}.
\end{itemize}

\begin{table}
	\centering
	\tiny
	\caption{The uncovered relations between detected communities (i.e., parties) by using DAAC in the Australia dataset.}
	\label{tab:australia_relations}
	\begin{tabular}{|m{1cm}|m{1cm}|m{1cm}|m{1cm}|m{1cm}|m{1cm}|@{}m{0cm}@{}}
		\hhline{~|*6{-}|}
		\multicolumn{1}{c|}{} & Liberals & Nationalists & Liberal Nationalists & Labors & Greens & \tabularnewline[2ex]  \hhline{|*6{-}|}\noalign{\vskip.2pt}
		
		Liberals & \cellcolor{green!40} \hspace{3mm}87 & \cellcolor{green!40} \hspace{3mm}61 & \cellcolor{green!40} \hspace{3mm}34 & \cellcolor{red!40} \hspace{3mm}-21 & \cellcolor{red!40} \hspace{3mm}-32 & \tabularnewline[2ex] \hhline{|*6{-}|}\noalign{\vskip.2pt}
		
		Nationalists & \cellcolor{green!40} \hspace{3mm} 61 & \cellcolor{green!40} \hspace{3mm} 52 & \cellcolor{green!40} \hspace{3mm} 46 & \cellcolor{red!40} \hspace{3mm} -4 & \cellcolor{red!40} \hspace{3mm}-22 & \tabularnewline[2ex] \hhline{|*6{-}|}\noalign{\vskip.2pt}
		
		Liberal Nationalists & \cellcolor{green!40} \hspace{3mm} 34 & \cellcolor{green!40} \hspace{3mm} 46 & \cellcolor{green!40} \hspace{3mm} 39 & \cellcolor{red!40} \hspace{3mm} -4 & \cellcolor{red!40} \hspace{3mm} -61 & \tabularnewline[2ex] \hhline{|*6{-}|}\noalign{\vskip.2pt}
		
		Labors & \cellcolor{red!40} \hspace{3mm} -21 & \cellcolor{red!40} \hspace{3mm} -4 & \cellcolor{red!40} \hspace{3mm} -4 & \cellcolor{green!40} \hspace{3mm} 121 & \cellcolor{red!40} \hspace{3mm} -31 & \tabularnewline[2ex] \hhline{|*6{-}|}\noalign{\vskip.2pt}
		
		Greens & \cellcolor{red!40} \hspace{3mm} -32 & \cellcolor{red!40} \hspace{3mm} -22 & \cellcolor{red!40} \hspace{3mm} -61 & \cellcolor{red!40} \hspace{3mm} -31 & \cellcolor{green!40} \hspace{3mm} 64 & \tabularnewline[2ex] \hhline{|*6{-}|}\noalign{\vskip.2pt}
	\end{tabular}
	\emph{Note:} all values in the table are rounded.
\end{table}

\subsection{Evaluation of Inter-community Relations}
In this section, we evaluate the effectiveness of our proposed framework in uncovering inter-community relations by conducting two experiments. To the best of our knowledge, there is no previous work to discover inter-community antagonistic and allied relations. Therefore, as the first experiment, we compare the inter-community relations which our framework detects with the real-word inter-community relations. Each community detected by our framework is labeled with the party to which the majority of its members belong. Then, we evaluate inter-community relations (i.e., matrix $\mathbf{H}$) detected by our algorithm according to the known ground-truth inter-party relations as previously presented in Section \ref{data-desc}.



Table \ref{tab:us_relations} shows intra/inter-community relation matrix $\mathbf{H}$ for the US dataset as well as the parties corresponding to the detected communities. In 2016, the Republican Party and the Democratic Party were strongly antagonistic towards each other, especially due to the 2016 presidential election campaigning\footnote{\url{https://en.wikipedia.org/wiki/United_States_presidential_election,_2016}} \cite{lilleker2016us}. As Table \ref{tab:us_relations} shows, our framework uncovers the existence of strong antagonism between these two parties. It also discovers that intra-community attitudes among the members of each community are highly positive as expected owing to the election campaign dynamics.

Table \ref{tab:australia_relations} shows intra/inter-community relation matrix $\mathbf{H}$ for the Australia dataset as well as the parties corresponding to the detected communities. The Liberal Party, the National Party, and the Liberal National party forged a coalition in the 2016 federal election. Except the relations between the members of the coalition, other relations among all parties were antagonistic\footnote{\url{https://en.wikipedia.org/wiki/Australian_federal_election,_2016}}. As shown in Table \ref{tab:australia_relations}, our framework uncovers the coalition in which the three involved parties are in alliance with each other. It also discovers antagonism between the members of the coalition and other parties as well as the antagonism between the Greens and the Labor Party. Moreover, it detects high positive intra-community attitudes among the members of communities as expected.

Table \ref{tab:uk_relations} shows intra/inter-community relation matrix $\mathbf{H}$ for the UK dataset as well as the parties corresponding to the detected communities. In 2015, there were antagonisms between all five major UK political parties, especially due to the 2015 general election campaigning\footnote{\url{https://en.wikipedia.org/wiki/United_Kingdom_general_election,_2015}} \cite{moran2015politics}. As shown in Table \ref{tab:uk_relations}, our framework correctly detects all antagonistic relations between these parties. It also discovers that intra-community attitudes among the members of each community are highly positive as expected.

\begin{table}
	\centering
	\tiny
	\caption{The uncovered relations between detected communities (i.e., parties) by using DAAC in the UK dataset.}
	\label{tab:uk_relations}
	\begin{tabular}{|m{1cm}|m{1cm}|m{1cm}|m{1cm}|m{1cm}|m{1cm}|@{}m{0cm}@{}}
		\hhline{~|*6{-}|}
		\multicolumn{1}{c|}{} & Conservatives & Labours & Lib dems & SNPs & UKIPs & \tabularnewline[2ex]  \hhline{|*6{-}|}\noalign{\vskip.2pt}
		
		Conservatives & \cellcolor{green!40} \hspace{3mm} 154 & \cellcolor{red!40} \hspace{3mm} -37 & \cellcolor{red!40} \hspace{3mm} -7 & \cellcolor{red!40} \hspace{3mm} -21 & \cellcolor{red!40} \hspace{3mm} -9 & \tabularnewline[2ex] \hhline{|*6{-}|}\noalign{\vskip.2pt}
		
		Labours & \cellcolor{red!40} \hspace{3mm} -37 & \cellcolor{green!40} \hspace{3mm} 242 & \cellcolor{red!40} \hspace{3mm} -8 & \cellcolor{red!40} \hspace{3mm} -11 & \cellcolor{red!40} \hspace{3mm} -26 & \tabularnewline[2ex] \hhline{|*6{-}|}\noalign{\vskip.2pt}
		
		Lib dems & \cellcolor{red!40} \hspace{3mm} -7 & \cellcolor{red!40} \hspace{3mm} -8 & \cellcolor{green!40} \hspace{3mm} 63 & \cellcolor{red!40} \hspace{3mm} -3 & \cellcolor{red!40} \hspace{3mm} -14 & \tabularnewline[2ex] \hhline{|*6{-}|}\noalign{\vskip.2pt}
		
		SNPs & \cellcolor{red!40} \hspace{3mm} -21 & \cellcolor{red!40} \hspace{3mm} -11 & \cellcolor{red!40} \hspace{3mm} -3 & \cellcolor{green!40} \hspace{3mm} 55 & \cellcolor{red!40} \hspace{3mm} -5 & \tabularnewline[2ex] \hhline{|*6{-}|}\noalign{\vskip.2pt}
		
		UKIPs & \cellcolor{red!40} \hspace{3mm} -9 & \cellcolor{red!40} \hspace{3mm} -26 & \cellcolor{red!40} \hspace{3mm} -14 & \cellcolor{red!40} \hspace{3mm} -5 & \cellcolor{green!40} \hspace{3mm} 30 & \tabularnewline[2ex] \hhline{|*6{-}|}\noalign{\vskip.2pt}
	\end{tabular}
	\emph{Note:} all values in the table are rounded.
\end{table}

The second experiment compares our framework with a two-step approach described as follows. We first utilize social interactions to detect communities. Then, we aggregate the sentiment expressed among the members of different communities in order to figure out their inter-community relations. To have a fair comparison, we use Eq. \eqref{proposed_graph-reg} to detect communities for the two-step approach; which is the main component in DAAC for utilizing social interactions. As Table \ref{tab:inter-community-accuracy} shows, the two-step approach is able to detect correct relations in US and Australia datasets. However, it fails to detect two out of ten inter-community relations in the UK dataset. This result shows that our proposed framework can detect inter-community relations more accurately by jointly using and social interactions and attitudes among users compared to a approach which sequentially detects communities and their relations.

\bgroup
\def\arraystretch{1.5}%
\begin{table}
	\centering
	\caption{Inter-community detection performance between DAAC and the two-step approach.}
	\label{tab:inter-community-accuracy}
	\begin{tabular}{|l |*{3}{>{\centering\arraybackslash}p{.15\linewidth}|}}
		\cline{2-4}
		\multicolumn{1}{c|}{}& \textbf{US} & \textbf{Australia} & \textbf{UK} \\ \hline
		Two-step approach & $1.0$ & $1.0$ & $0.8$ \\ \hline
		DAAC & $1.0$ & $1.0$ & $1.0$ \\ \hline
	\end{tabular}
\end{table}
\egroup

\pgfplotsset{
	small,
	legend style={legend pos=south east},
	xlabel=Regularization parameter,
	y label style={at={(axis description cs:0.1,.5)}},
	ylabel=Performance,
}%

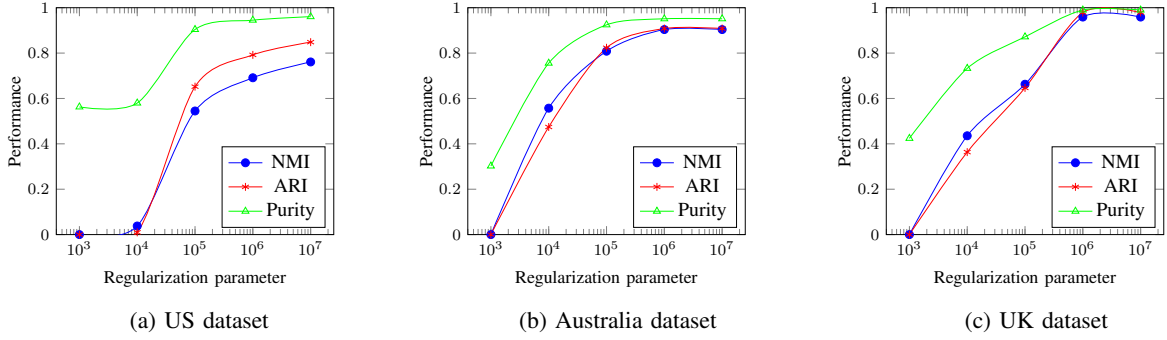
\begin{figure*}[!htb]
	\centering
	\begin{minipage}[b]{0.3\textwidth}
		\scalebox{0.75}{
			\begin{tikzpicture}
			\begin{semilogxaxis}[
			ymin=0.0, ymax=1,
			]]    
			\addplot[smooth,mark=*,blue] plot coordinates {
				(1000,0.0)
				(10000,0.037854)
				(100000,0.544826)
				(1000000,0.691229)
				(10000000,0.760947)
				
			};
			\addlegendentry{NMI}
			
			\addplot[smooth,color=red,mark=asterisk]
			plot coordinates {
				(1000,0.0)
				(10000,0.009865)
				(100000,0.652069)
				(1000000,0.792109)
				(10000000,0.848143)
			};
			\addlegendentry{ARI}
			
			\addplot[smooth,color=green,mark=triangle]
			plot coordinates {
				(1000,0.562607)
				(10000,0.579760)
				(100000,0.903945)
				(1000000,0.945111)
				(10000000,0.960549)
			};
			\addlegendentry{Purity}    
			
			\end{semilogxaxis}
			\end{tikzpicture}  
		}
		\subcaption{US dataset}
	\end{minipage}
	\begin{minipage}[b]{0.3\textwidth}
		\scalebox{0.75}{\begin{tikzpicture}
			\begin{semilogxaxis}[
			ymin=0.0, ymax=1,
			]]    
			\addplot[smooth,mark=*,blue] plot coordinates {
				(1000,0.0)
				(10000,0.557212)
				(100000,0.808268)
				(1000000,0.903691)
				(10000000,0.903691)
				
			};
			\addlegendentry{NMI}
			
			\addplot[smooth,color=red,mark=asterisk]
			plot coordinates {
				(1000,0.0)
				(10000,0.474295)
				(100000,0.823392)
				(1000000,0.908264)
				(10000000,0.908264)
			};
			\addlegendentry{ARI}
			
			\addplot[smooth,color=green,mark=triangle]
			plot coordinates {
				(1000,0.302222)
				(10000,0.755556)
				(100000,0.924444)
				(1000000,0.951111)
				(10000000,0.951111)
			};
			\addlegendentry{Purity}    
			
			\end{semilogxaxis}
			\end{tikzpicture}  
		}
		\subcaption{Australia dataset}
	\end{minipage}
	\begin{minipage}[b]{0.3\textwidth}
		\scalebox{0.75}{\begin{tikzpicture}
			\begin{semilogxaxis}[
			ymin=0.0, ymax=1,
			]]    
			\addplot[smooth,mark=*,blue] plot coordinates {
				(1000,0.0)
				(10000,0.435580)
				(100000,0.662663)
				(1000000,0.958806)
				(10000000,0.958806)
				
			};
			\addlegendentry{NMI}
			
			\addplot[smooth,color=red,mark=asterisk]
			plot coordinates {
				(1000,0.0)
				(10000,0.363539)
				(100000,0.646499)
				(1000000,0.978770)
				(10000000,0.978770)
			};
			\addlegendentry{ARI}
			
			\addplot[smooth,color=green,mark=triangle]
			plot coordinates {
				(1000,0.424165)
				(10000,0.732648)
				(100000,0.871465)
				(1000000,0.989717)
				(10000000,0.989717)
			};
			\addlegendentry{Purity}    
			
			\end{semilogxaxis}
			\end{tikzpicture}  
		}
		\subcaption{UK dataset}
	\end{minipage}
	\caption{Community detection performance with regard to $\lambda$.}
	\label{fig:1}
\end{figure*}


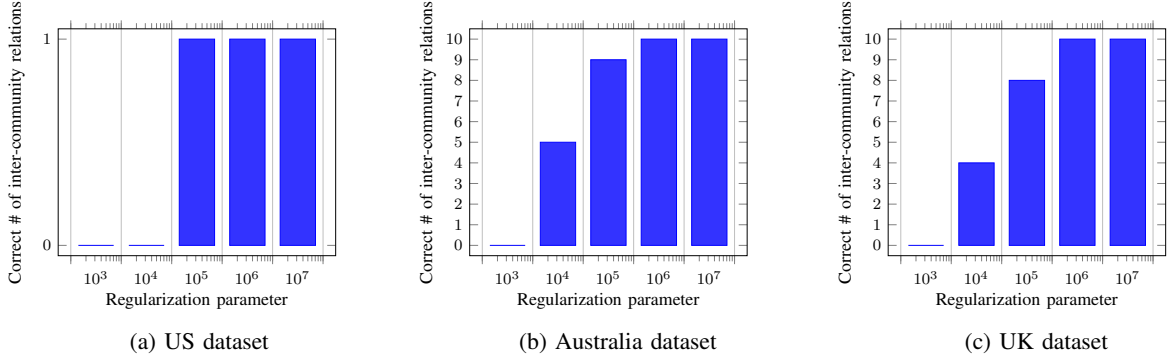
\begin{figure*}[!htb]
	\centering
	\begin{minipage}[b]{0.3\textwidth}
		\scalebox{0.75}{
			\begin{tikzpicture}
			\begin{semilogxaxis}[
			ylabel=Correct \# of inter-community relations,
			ymin=0, ymax=1,
			ytick={0,1},
			enlargelimits=0.05,
			ybar interval=0.7
			]    
			\addplot[blue!20!blue,fill=blue!80] plot coordinates {
				(1000,0)
				(10000,0)
				(100000,1)
				(1000000,1)
				(10000000,1)
				(100000000,1)
			};
			\end{semilogxaxis}
			\end{tikzpicture}
		}
		\subcaption{US dataset}
	\end{minipage}
	\begin{minipage}[b]{0.3\textwidth}
		\scalebox{0.75}{\begin{tikzpicture}
			\begin{semilogxaxis}[
			ymin=0, ymax=10,
			ylabel=Correct \# of inter-community relations,
			ytick={0,1,2,3,4,5,6,7,8,9,10},
			enlargelimits=0.05,
			ybar interval=0.7,
			]    
			\addplot[blue!20!blue,fill=blue!80] plot coordinates {
				(1000,0)
				(10000,5)
				(100000,9)
				(1000000,10)
				(10000000,10)
				(100000000,10)
			};
			\end{semilogxaxis}
			\end{tikzpicture}
		}
		\subcaption{Australia dataset}
	\end{minipage}
	\begin{minipage}[b]{0.3\textwidth}
		\scalebox{0.75}{\begin{tikzpicture}
			\begin{semilogxaxis}[
			ymin=0, ymax=10,
			ylabel=Correct \# of inter-community relations,
			ytick={0,1,2,3,4,5,6,7,8,9,10},
			enlargelimits=0.05,
			ybar interval=0.7
			]    
			\addplot[blue!20!blue,fill=blue!80] plot coordinates {
				(1000,0)
				(10000,4)
				(100000,8)
				(1000000,10)
				(10000000,10)
				(100000000,10)
			};
			\end{semilogxaxis}
			\end{tikzpicture}
		}
		\subcaption{UK dataset}
	\end{minipage} 
	\caption{The correct number of inter-community relations with regard to $\lambda$.}
	\label{fig:2}
\end{figure*}

\subsection{Study on the Regularization Parameter}
\label{reg-parameter}
In this section, we investigate the sensitivity of our framework with respect to regularization parameter $\lambda$. We vary the value of $\lambda$, and plot NMI, ARI and Purity measures in Figure \ref{fig:1} for all three datasets used in the study. Similarly, we plot the correct number of inter-community relations discovered by DAAC in Figure \ref{fig:2} for all three datasets with respect to different values of $\lambda$.

As we observe from Figure \ref{fig:1}, very large values of $\lambda$ (e.g., $10^6$ and $10^7$) for all datasets result in the highest performance of DAAC in detecting communities. Similarly, Figure \ref{fig:2} shows that very large values of $\lambda$ also result in the highest number of correct inter-community relations discovered by DAAC. The rationale behind this is that inter-community relations cannot be correctly identified unless communities are accurately detected.

\section{Conclusion and Future Work}
\label{conclusion}
In this paper, we proposed a framework to discover communities and their relations by exploiting social interactions and user-generated content. We validated the hypothesis that inter-community attitudes that users express towards each other in social media can reflect inter-community relations. As inspired by this hypothesis, our proposed framework DAAC jointly models users' attitudes and social interactions in order to uncover communities and their antagonistic/allied relations. Experimental results on three real-world social media datasets demonstrated that our framework obtains significant performance in detecting communities compared with several baselines and also detects inter-community relations correctly. Moreover, we showed that a two-step approach, which sequentially detect communities and their relations, can fail to detect correct inter-community relations. 

Since communities and their relations evolve over time, studying such dynamics provides deeper insights into understanding communities. In our future work, we aim to study uncovering the dynamics of communities and their relations and the motives behind these dynamics.

\appendix
\section{Appendix}
Optimizing the objective function $\bm{\mathcal{F}}$ in Eq. \eqref{proposed_framework} with respect to $U$ is equivalent to solving

\begin{equation}
\begin{aligned}
\underset{\mathbf{U}}{\text{min}}\quad & \bm{\mathcal{F}}_\mathbf{U} = ||\mathbf{S} - \mathbf{U} \mathbf{H} \mathbf{U}^T ||_F^2 - \lambda Tr(\mathbf{U}^T \overset{\sim}{\mathbf{R}} \mathbf{U})\\
\text{s.t.}\quad & \mathbf{U} \geq 0, \mathbf{U}^T \mathbf{U} = \mathbf{I}.
\end{aligned}
\end{equation}

Let $\mathbf{\Gamma}$ and $\mathbf{\Lambda}$ be the Lagrange multiplier for constraints $\mathbf{U}^T \mathbf{U} = \mathbf{I}$ and $\mathbf{U} \geq 0$, respectively, and the Lagrange function is defined as follows:
\begin{equation}
\begin{aligned}
\underset{\mathbf{U}}{\text{min}} \quad & \bm{\mathcal{L}}_\mathbf{U} = ||\mathbf{S} - \mathbf{U} \mathbf{H} \mathbf{U}^T ||_F^2 - \lambda Tr(\mathbf{U}^T \overset{\sim}{\mathbf{R}} \mathbf{U})\\
& \quad - Tr(\mathbf{\Lambda} \mathbf{U}^T) +  Tr(\mathbf{\Gamma} (\mathbf{U}^T \mathbf{U} - \mathbf{I})) \\
\end{aligned}
\end{equation}

The derivative of $\bm{\mathcal{L}}_\mathbf{U}$ with respect to $\mathbf{U}$ is
\begin{equation}
\begin{aligned}
&\frac{\partial \bm{\mathcal{L}}_\mathbf{U}}{\partial \mathbf{U}} = 
-2(\mathbf{W} \odot \mathbf{W} \odot \mathbf{S}) \mathbf{U} \mathbf{H}^T -2(\mathbf{W} \odot \mathbf{W} \odot \mathbf{S})^T \mathbf{U} \mathbf{H}\\
& \quad + 2(\mathbf{W} \odot \mathbf{W} \odot \mathbf{U} \mathbf{H} \mathbf{U}^T) \mathbf{U} \mathbf{H}^T \\
& \quad + 2(\mathbf{W} \odot \mathbf{W} \odot \mathbf{U} \mathbf{H} \mathbf{U}^T)^T \mathbf{U} \mathbf{H}\\
& \quad - 2 \lambda \overset{\sim}{\mathbf{R}} \mathbf{U} - \mathbf{\Lambda} + 2 \mathbf{U} \mathbf{\Gamma}
\end{aligned}
\end{equation}

For the sake of simplicity, let us assume that,
\begin{align}
& \mathbf{E}_1 = -(\mathbf{W} \odot \mathbf{W} \odot \mathbf{S}) \mathbf{U} \mathbf{H}^T  \\
& \mathbf{E}_2 = -(\mathbf{W} \odot \mathbf{W} \odot \mathbf{S})^T \mathbf{U} \mathbf{H} \\
& \mathbf{E}_3 = (\mathbf{W} \odot \mathbf{W} \odot \mathbf{U} \mathbf{H} \mathbf{U}^T) \mathbf{U} \mathbf{H}^T \\
& \mathbf{E}_4 = (\mathbf{W} \odot \mathbf{W} \odot \mathbf{U} \mathbf{H} \mathbf{U}^T)^T \mathbf{U} \mathbf{H}
\end{align}

By setting $\frac{\partial \bm{\mathcal{L}}_\mathbf{U}}{\partial \mathbf{U}} = 0$, we get
\begin{equation}
\Lambda = -2\mathbf{E}_1 -2\mathbf{E}_2 + 2\mathbf{E}_3 + 2\mathbf{E}_4 - 2 \lambda \overset{\sim}{\mathbf{R}} \mathbf{U} + 2 \mathbf{U} \mathbf{\Gamma}
\end{equation}

With the KKT complementary condition for the nonnegativity of $\mathbf{U}$, we have
\begin{equation}
\mathbf{\Lambda}_{ij} \mathbf{U}_{ij} = 0
\end{equation}

Therefore, we have
\begin{equation}
(\mathbf{E}_1 + \mathbf{E}_2 + \mathbf{E}_3 + \mathbf{E}_4 - \lambda \overset{\sim}{\mathbf{R}} \mathbf{U} + 2 \mathbf{U}\mathbf{\Gamma}) _{ij} \mathbf{U}_{ij} = 0
\end{equation}
where
\begin{equation}
\mathbf{\Gamma} = - \mathbf{U}^T \mathbf{E}_1 - \mathbf{U}^T \mathbf{E}_2 - \mathbf{U}^T \mathbf{E}_3 - \mathbf{U}^T \mathbf{E}_4 + \lambda \mathbf{U}^T \overset{\sim}{\mathbf{R}} \mathbf{U}
\end{equation}

Since $\mathbf{E}_1$, $\mathbf{E}_2$, $\mathbf{E}_3$, $\mathbf{E}_4$, and $\mathbf{\Gamma}$ can take mixed signs. Suggested by \cite{ding2010convex}, we separate positive and negative parts of any matrix $A$ as
\begin{equation}
\begin{aligned}
& \mathbf{A}_{ij}^+ = (|\mathbf{A}_{ij}| + \mathbf{A}_{ij})/2 \\ & \mathbf{A}_{ij}^- = (|\mathbf{A}_{ij}| - \mathbf{A}_{ij})/2
\end{aligned}
\end{equation}

Then, we get the following update rule of $\mathbf{U}$,
\begin{equation}
\mathbf{U} = \mathbf{U} \odot \sqrt{\frac{\mathbf{E}_1^+ + \mathbf{E}_2^+ + \mathbf{E}_3^- + \mathbf{E}_4^- + \lambda \overset{\sim}{\mathbf{R}} \mathbf{U} + \mathbf{U}\mathbf{\Gamma}^-} {\mathbf{E}_1^- + \mathbf{E}_2^- + \mathbf{E}_3^+ + \mathbf{E}_4^+ + \mathbf{U} \mathbf{\Gamma}^+}}
\end{equation}

\section*{Acknowledgments}
This work was partially supported by ONR Grant N00014-16-1-2015 and USAF Grant FA9550-15-1-0004.

\bibliography{references}
\bibliographystyle{IEEEtran}



\end{document}